\documentclass[twocolumn,prd,aps,superscriptaddress,preprintnumbers,tightenlines,showpacs,nofootinbib,eqsecnum,amsfonts,amsmath]{revtex4-1}
\usepackage{placeins}
\usepackage{color}
\usepackage{calc}
\usepackage{amsmath,amssymb,graphicx}
\usepackage{tensor}
\usepackage{bm}
\usepackage{times}
\usepackage[varg]{txfonts}
\usepackage[colorlinks, pdfborder={0 0 0}]{hyperref}
\usepackage{float}
\usepackage{dcolumn}
\usepackage[nolist,nohyperlinks]{acronym}
\usepackage{xspace}
\usepackage[abs]{overpic}
\usepackage{pict2e}
\usepackage{enumitem}
\usepackage[usenames,dvipsnames]{xcolor}
\usepackage[utf8]{inputenc}
\usepackage{acronym}
\usepackage{gensymb}
\usepackage{cleveref}
\usepackage[normalem]{ulem}
\usepackage{longtable}
\usepackage{multirow}

{\renewcommand{\arraystretch}{1.5}
\newcommand{\chirpm}{\mathcal{M}_c}

\begin{document}
 
 \title{Tests of general relativity with stellar-mass black hole binaries observed by LISA}

\author{Alexandre Toubiana}
\affiliation{APC, AstroParticule et Cosmologie, Universit\'e Paris Diderot,
CNRS/IN2P3, CEA/Irfu, Observatoire de Paris, Sorbonne Paris Cit\'e,
10, rue Alice Domon et L\'eonie Duquet 75205 PARIS Cedex 13, France}
\affiliation{Institut d'Astrophysique de Paris, CNRS \& Sorbonne
 Universit\'es, UMR 7095, 98 bis bd Arago, 75014 Paris, France} 

\author{Sylvain Marsat}
\affiliation{APC, AstroParticule et Cosmologie, Universit\'e Paris Diderot,
CNRS/IN2P3, CEA/Irfu, Observatoire de Paris, Sorbonne Paris Cit\'e,
10, rue Alice Domon et L\'eonie Duquet 75205 PARIS Cedex 13, France}

\author{Stanislav Babak}
\affiliation{APC, AstroParticule et Cosmologie, Universit\'e Paris Diderot,
CNRS/IN2P3, CEA/Irfu, Observatoire de Paris, Sorbonne Paris Cit\'e,
10, rue Alice Domon et L\'eonie Duquet 75205 PARIS Cedex 13, France}
\affiliation{Moscow Institute of Physics and Technology, Dolgoprudny, Moscow region, Russia}

\author{Enrico Barausse}
\affiliation{SISSA, Via Bonomea 265, 34136 Trieste, Italy and INFN Sezione di Trieste, Via Valerio 2, 34127 Trieste, Italy}
\affiliation{IFPU - Institute for Fundamental Physics of the Universe, Via Beirut 2, 34014 Trieste, Italy}
\affiliation{Institut d'Astrophysique de Paris, CNRS \& Sorbonne
 Universit\'es, UMR 7095, 98 bis bd Arago, 75014 Paris, France}

\author{John Baker}
\affiliation{Gravitational Astrophysics Laboratory, NASA Goddard Space Flight Center, 8800 Greenbelt Road., Greenbelt, Maryland 20771, USA}

 \begin{abstract}
We consider the observation of stellar-mass black holes binaries with the Laser Interferometer Space Antenna (LISA). Preliminary  results based on Fisher information matrix analyses
have suggested that gravitational waves from those sources could be very sensitive to possible deviations from the theory of general relativity and from the strong equivalence principle during the low-frequency binary inspiral. We perform a full Markov Chain Monte Carlo Bayesian analysis to quantify the
sensitivity of these signals to two phenomenological modifications of general relativity, namely a
putative gravitational dipole emission and a nonzero mass for the graviton, properly accounting for the detector's response. Moreover, we consider a scenario where those sources could be observed also with Earth-based detectors, which should measure the coalescence time with precision better than  $1 \ {\rm ms}$. This constraint on the coalescence time further improves the bounds that we can set on those phenomenological deviations from general relativity.
We show that tests of dipole radiation and the graviton's mass should improve respectively by seven and half an order(s)  of magnitude over current bounds. Finally, we discuss under which conditions one may claim the detection of a modification to general relativity. 
 \end{abstract}
    
 \maketitle
 
\section{Introduction}

The first detections by the LIGO/VIRGO collaboration have shown the potential of gravitational waves (GWs) to explore the universe and to investigate the fundamental laws of physics. These observations have provided evidence for the existence of an astrophysical population of black hole binaries and are also in a very good agreement with the predictions of Einstein's theory of general relativity (GR)~\cite{Abbott:2016blz,Abbott:2017oio,Abbott:2017gyy,TheLIGOScientific:2016pea,LIGOScientific:2018mvr,LIGOScientific:2019fpa,TheLIGOScientific:2016src}.
These black holes have larger masses than originally expected by most of the community (see however \cite{Spera:2015vkd}), up to $\sim 50 M{\odot}$~\cite{LIGOScientific:2018mvr}.
Such relatively heavy  stellar-mass black hole binaries (SBHBs)\footnote{In SBHBs we also include possible primordial black holes.} could in principle be observed also by the space-based Laser Interferometer Space Antenna (LISA)~\cite{Sesana:2016ljz, Moore:2019pke}. Scheduled for launch in 2034, LISA \cite{AHO17} will be sensitive to lower frequencies (mHz) than terrestrial detectors such as LIGO and VIRGO. Massive and supermassive black hole binaries with total masses $M\sim10^4 -10^7 M{\odot}$  will be its primary target \cite{Klein:2015hvg},
but the observation of the early inspiral of SBHBs will be complementary to the operations of
ground based detectors. The next generation of ground-based interferometers will indeed observe the mergers of these sources, after they have left the LISA band and reemerged, typically a few weeks, months or even years later, in their higher frequency band~\cite{Sesana:2016ljz,AmaroSeoane:2009ui}.

Since LISA will observe SBHBs in the long low-frequency inspiral phase, which for these sources can last  for years,
the accuracy and precision with which we will recover the intrinsic and extrinsic source parameters
are expected to improve over what could be achieved with ground interferometers alone~\cite{Sesana:2016ljz,Gnocchi:2019jzp,Carson:2019rda,Vitale:2016rfr,Barausse:2016eii,Chamberlain:2017fjl,Toubiana:2020cqv}. Furthermore, multiband observations of these SBHBs by LISA and terrestrial detectors will track the evolution of the GW signal across several orders of magnitudes in frequency, and provide us with even more accurate determinations of the parameters.

Tracking the phase of these binaries for extended periods of  time should allow for the detection of low-frequency modifications
of the waveform due to the interaction with matter and to the system's peculiar acceleration~\cite{Barausse:2014tra,Barausse:2014pra,Tamanini:2019usx,Caputo:2020irr}. In addition, these observations will allow us to perform exquisite tests of the strong equivalence principle. More specifically, they should permit 
studying the possible presence of black hole hairs and extra polarisation states via
their backreaction on the orbital evolution, and testing nonlinear dispersion relations~\cite{Barausse:2016eii}. Indeed, the observation of SBHBs with LISA, alone or in joint operations with ground detectors, is expected to improve current bounds on these effects by several orders of magnitude.

The aim of this paper is to assess how well LISA can constrain deviations from GR, and detect them if they are present in SBHBs signals. In addition we will assess  how much improvement we could expect from multiband observations. 
 Previous works on this topic~\cite{Sesana:2016ljz,Gnocchi:2019jzp,Carson:2019rda,Vitale:2016rfr,Barausse:2016eii, Chamberlain:2017fjl,Carson:2019kkh}
used Fisher information matrices to perform parameter estimation. However, this technique, while  quick and efficient, is usually not suited for  (i) events with low signal-to-noise ratio (SNR) \cite{Vallisneri:2007ev} and for (ii) non-Gaussian and/or multimodal distributions. Furthermore, since SBHBs can stay in the
high-frequency band ($\sim  10^{-2}$ Hz) of the  LISA sensitivity for years (and in some cases for 
the whole mission duration), the response of the detector has to be properly taken into account. In particular, the use of the long wavelength approximation for the response could impact the parameter estimation results~\cite{Vecchio:2004vt,Vecchio:2004ec}. To address these issues and improve on existing results, we therefore account for the full LISA response function and we perform a full Bayesian analysis of the parameter estimation
of SBHBs with LISA, for a number of fiducial representative systems. Our Bayesian treatment also takes a step toward understanding how LISA data might be applied for this kind of study in practice.

Our analysis shows that LISA's observations alone will improve tests of GR and of the equivalence principle 
(namely tests of dipole radiation and the graviton's mass)
by respectively seven and half an order(s) of magnitude over the current bounds. In addition, if deviations from GR are significant, we should be able to confidently detect them, even if the deviations are below current bounds. Finally, we show that multiband observations should enable us to break  degeneracies among the parameters and further improve detectability of (or bounds on) possible deviations from GR.

This paper is organised as follows. In Sec. \ref{ppe_gw} we present the deviations from GR that we consider, and their effect on the GW signal. Then we present the details of our method, and in particular how data are simulated and how we perform the Bayesian analysis, in Sec. \ref{meth}. 
Our results are presented in Sec. \ref{res} and our conclusions in Sec. \ref{ccl}.

\section{Parametric deviations from General Relativity}\label{ppe_gw}

Deviations from GR can potentially affect both the generation of GWs and their propagation to the detector.
In this paper, we will focus on two specific examples, dipolar GW emission and a nonzero mass for the graviton.

GW emission in GR starts at the quadrupolar order, while no monopole or dipole gravitational emission is present. This happens
because of the covariant conservation of the matter stress energy tensor \cite{Will2014}, which leads to conservation of energy and linear momentum, just as monopole emission  
in electromagnetism is forbidden because the Maxwell equations imply the conservation of the electric charge.

However, in theories of gravity that modify/extend GR, extra gravitational fields (besides
the spin-2 graviton) are typically present, see e.g. \cite{Berti:2015itd} for a review. These extra fields, while not coupled  directly to matter  in order to
enforce the validity of weak equivalence principle (i.e. the universality of free fall), are typically coupled nonminimally to the spin-2 field. 
Since the spin-2 graviton couples to matter, the latter is effectively coupled also to the extra graviton fields via the spin-2 metric perturbations. This spin-2 mediated interaction is negligible when the metric perturbations are small (i.e. when gravity is weak), but can
become important when the gravitational field is strong.
This effect (often referred to as ``Nordtvedt effect'' \cite{Eardley75,Nordtvedt:1969zz,Roll:1964rd} takes place in the interior/vicinity of compact objects such as neutron stars and black holes. As a result, compact objects may experience an effective coupling to the extra gravitational fields, 
which can give rise to ``fifth forces''. These forces will depend in general on the nature and composition of the compact object, and vanish 
in the limit where the object's compactness is low (i.e. they vanish when the self gravity of the body is weak). Therefore, they cause
violations of the universality of free fall. However, since these effects only take place for strongly gravitating objects, they are often referred to as 
violations of the ``strong equivalence principle''.

Violations of the strong equivalence principle can be both dissipative (i.e. affecting the GW fluxes) or conservative [i.e. modifying the Newtonian interactions
of compact objects and their post-Newtonian (PN)\footnote{A correction is referred to as PN correction of order $n$ (nPN) if it
is of order $(v/c)^{2n}$ relative to the leading order term, where $v$ is the characteristic velocity of the system.} conservative corrections]. It was discovered early on, for instance, that in scalar-tensor theories of the Fierz-Jordan-Brans-Dicke type \cite{Fierz:1956zz,Jordan:1959eg,Brans:1961sx}, the dynamics of quasicircular neutron star binaries is modified (with respect to GR) by the appearance  of dipole (-1PN) gravitational fluxes (which can be interpreted 
 as exchanges of energy and momentum from the binary to the gravitational scalar, due to the Nordtvedt effect), 
 and by conservative corrections to the Newtonian and PN interaction of the two bodies \cite{Damour:1992we,Will:1989sk,Alsing:2011er}. Note that monopole GW emission is also possible in principle in these scalar-tensor theories, but it is suppressed in quasicircular systems \cite{Damour:1992we,Will:1989sk,Alsing:2011er}. 

Dipole emission has a strong effect on the binary evolution as it increases the rate of change of the orbital frequency, i.e. dipole GW emission, being a -1PN effect, is potentially more important than the quadrupole GW flux. As a result, 
the binary loses more energy to GWs at low frequencies, which, in turn, translates into a faster orbital evolution. Indeed, the absence of dipole GW emission in the dynamics of known binary pulsar systems (whose evolution is tracked by radio observations \cite{Damour:1991rd})
has allowed for placing stringent constraints on scalar-tensor theories of the Fierz-Jordan-Brans-Dicke type \cite{Damour:1993hw,Damour:1996ke,Freire:2012mg}. For black hole binaries, on the other hand, these scalar-tensor theories predict that dynamics should
be unaffected by the Nordtvedt effect (i.e. no deviations from GR should be present in both the conservative and dissipative dynamics), at least if their 
spacetime is asymptotically flat \cite{Healy:2011ef,Horbatsch:2011ye,Berti:2013gfa}.

More recently, however, it has been recognised that  binary black holes may also experience violations of the strong equivalence principle in theories of gravity that extend GR. 
In fact, even in Fierz-Jordan-Brans-Dicke scalar-tensor theories, black holes may acquire a ``hair'', i.e. a coupling to the gravitational scalar, if their spacetime is not asymptotically flat 
(e.g. due to cosmological boundary conditions or the presence of nearby matter) \cite{Healy:2011ef,Horbatsch:2011ye,Berti:2013gfa}. These hairs would then produce GW dipole emission and modifications to the conservative dynamics of binary black hole systems. 
Similar black hole hairs are naturally produced also in more generic scalar-tensor theories (e.g. Horndeski theories, dilaton-Maxwell theories) \cite{2014PhRvL.112y1102S,Yagi:2015oca,Barausse:2015wia,Silva:2017uqg,Herdeiro:2018wub,Julie:2018lfp,Julie:2019sab}, 
where they can even become significantly large (``nonperturbative'') in specific situations \cite{Silva:2017uqg,Herdeiro:2018wub}.
 Black hole hairs are probably present also in theories with extra gravitational vector and tensor fields 
 (e.g. in some regions of the parameter space of Lorentz violating gravity \cite{Ramos:2018oku}, in massive gravity \cite{Babichev:2015xha}, etc.).

GW observations can be used  to experimentally test the possibility that the Nordtvedt effect may be at play in black hole binaries. Observations of SBHBs with LISA will be ideal to this purpose, because they will 
probe the low-frequency evolution of these systems, where dipole emission (being a -1PN effect) could potentially dominate over the GR evolution. Indeed, \cite{Barausse:2016eii}  
used a Fisher information matrix analysis to suggest that LISA observations of these systems 
(or even better, joint observations by LISA and ground based detectors) could test the presence of vacuum dipole GW emission to a precision challenging the the one achieved with binary pulsar systems. In this paper, we will 
extend that work by employing more rigorous Bayesian techniques and by accounting for the full response of the LISA detector, which was not considered in \cite{Barausse:2016eii}, in spite of being crucial for SBHBs.

A theory-agnostic phenomenological framework to describe and classify deviations from GR, at least during the low frequency inspiral, is provided by the parameterised post-Einsteinian (ppE)
formalism \cite{Yunes:2009ke}. A similar formalism is applied in the TIGER pipeline \cite{Agathos:2013upa} used in LIGO/VIRGO tests of GR \cite{TheLIGOScientific:2016src}.
In these approaches the general relativistic phase and amplitude are modified as
\begin{equation}\label{PPE}
 \tilde{h}(f)=\tilde{h}_{GR}(1+\gamma (\pi \chirpm f)^c)e^{i\delta (\pi \chirpm f)^d}\,,
\end{equation}
where $\tilde{h}_{GR}$ is the frequency domain waveform of GR, while the deviations from GR are described by the dimensionless parameters $\gamma$, $\delta$, $c$ and $d$. In the expression above, $\chirpm=\left ( \frac{m_1^3m_2^3}{m_1+m_2}\right )^{1/5}$, is the chirp mass of the binary.
Since interferometers are mostly sensitive to the phase of GWs,  we will neglect the amplitude modifications and set  $\gamma=0$. This approximation has been discussed and justified in \cite{Tahura:2019dgr}, to which we refer for further details.
Different values of $\delta$ and $d$   
correspond  to distinct physical effects and gravitational theories (see e.g. \cite{Barausse:2016eii,Yagi:2011xp,Yagi:2015oca,Yunes:2009bv,Yunes:2010yf,Yagi:2011yu,Yunes:2011we,Vigeland:2011ji,Mirshekari:2011yq,Carson:2019kkh} for some specific modifications of GR and their mapping to the PPE parameters $\delta$ and $b$).

Following \cite{Barausse:2016eii} we parametrise the dipole GW energy flux as
 \begin{equation}\label{dip_rad}
  \dot{E}_{tot}=\dot{E}_{GR}(1+Bv^{-2}),
 \end{equation}
where $\dot{E}_{GR}$ is the GR quadrupole flux, $v$ is the relative velocity in the binary, and $B$ (which vanishes in GR) is a theory and system dependent parameter that characterises dipole emission. 
Given this modified energy loss rate, we can compute the frequency evolution of the system by using the stationary phase approximation \cite{Cutler:1994ys,Buonanno:2007yg,2014ITSP}. 
Assuming a small $B$,  we obtain that the ppE coefficients corresponding to dipole emission are \cite{Barausse:2016eii,Yunes:2016jcc}:
\begin{align}
 \delta&=-\frac{3}{224}\eta^{2/5}B\,, \\
 d&={-7/3}\,,
\end{align}
where $\eta=m_1 m_2/(m_1+m_2)^2$ (with $m_1$ and $m_2$ the individual masses) is the symmetric mass ratio. 
Because of the violation of the strong equivalence principle, the value of $B$ might depend on the nature of the system. To make it explicit that we are considering dipolar radiation in black hole systems, we will use the symbol $B_{{\rm BH}}$ for the rest of this paper. 

Besides modifying the generation of GWs, deviations from GR may also affect  wave propagation. Phenomenologically,
that can be encoded in a modified dispersion relation \cite{Will:1997bb,Mirshekari:2011yq}
\begin{equation}\label{disp_rel}
 E^2=p^2c^2+\mathbb{A}_{\alpha}p^{\alpha}c^{\alpha}\,,
\end{equation}
where $E$ and $p$ are the graviton's energy and linear momentum, while $\mathbb{A}$ and $\alpha$ are free parameters. 
For example, Ho\v{r}ava gravity  predicts the presence of terms with both $\alpha=4$ and $\alpha=6$ \cite{Horava:2009uw,Visser:2009fg},
while the case $\alpha=0$ corresponds to a massive graviton \cite{Fierz:1956zz}.

From Eq.~\eqref{disp_rel}, at first order in $\mathbb{A}$ the graviton's velocity reads:
\begin{equation}
 \frac{v_g^2}{c^2}=1-\mathbb{A}E^{\alpha-2}\,.
\end{equation}
By measuring the time delay between the GWs and light emitted by GW170817, it is possible to bound the fractional difference
between the speed of GWs and that of light to less than $10^{-15}$ \cite{Monitor:2017mdv}. 
 A modified dispersion relation also deforms the shape of the GW signal as it propagates, since each frequency travels
at a different (phase and group) speed \cite{Mirshekari:2011yq}. This allows for testing modified dispersion relations even in the absence of electromagnetic counterpart, as is expected to be the case for SBHBs. This is the technique by which the LIGO/VIRGO collaboration \cite{TheLIGOScientific:2016src,LIGOScientific:2019fpa} has obtained graviton mass bounds competitive with solar system observations \cite{Will:2018gku}. 

In more detail, Ref.~\cite{Mirshekari:2011yq} showed that a modified dispersion relation like Eq.~\eqref{disp_rel} changes the phase of the GW signal, and this modification corresponds to ppE coefficients \cite{Mirshekari:2011yq}:
\begin{align}
 \delta&=-\frac{\pi^{2-\alpha}}{1-\alpha}\frac{D_{\alpha}\chirpm^{1-\alpha}}{(1+Z)^{1-\alpha}}\frac{\mathbb{A}}{(hc)^2}, \, \label{mod_disp}\\
 d&=\alpha -1\,,
\end{align}
where $D_{\alpha}$ is a distance variable given by $D_{\alpha}=\frac{(1+Z)^{1-\alpha}}{H_0}\int_0^Z \frac{(1+z')^{\alpha-2}}{\sqrt{\Omega_m(1+z')^3+\Omega_{\Lambda}}} {\rm d}z'$. In this expression, $Z$ is the cosmological redshift of the source, $H_{0}$ is the Hubble constant and $\Omega_m$ and $\Omega_{{\Lambda}}$ are the matter and dark energy density parameters respectively. For those last three quantities we take the values measured by the Planck mission \cite{Aghanim:2018eyx}.

In this paper we will focus on the case of a massive graviton, $\alpha=0$ and $\mathbb{A}=m_g^2 c^4$, where $m_g$ is the graviton's mass.
As can be seen in Eq.~\eqref{mod_disp}, the phase shift due to a massive graviton increases with chirp mass and distance. Thus, 
we may expect the best GW constraints on this effect to come not from GW170817, 
but from more massive and distant systems \cite{Abbott:2018lct}. For a given source, 
though, distance also reduces the SNR, so experimental bounds are defined by the interplay of those two factors.

\section{Method}\label{meth}

In this work we want to assess the sensitivity of LISA to modifications of GR. To that purpose, we work in a full Bayesian framework and consider two different Markov Chain Monte Carlo experiments. 
The first one consists of simulating the signal predicted by GR, for a few astrophysical systems, and trying to recover it with templates where either dipolar radiation or a massive graviton is allowed. Non-GR templates are computed within the framework presented in Sec. \ref{ppe_gw}.
Our goal is to place upper bounds on $B_{{\rm BH}}$ and $m_g$, i.e. to determine how well GW observations can constrain those deviations, accounting also for possible correlations between parameters. The second experiment consists of simulating signals containing a modification to GR, for the same astrophysical systems, and estimating  how well can we detect this modification.

In order to obtain an estimate of how much multiband observations could improve our ability to detect or constrain modifications to GR, we repeat each of the previous experiments, but placing a very tight prior constraint on the coalescence time. Indeed, this parameter is extremely well constrained by ground based detectors, with a typical accuracy of a few milliseconds. 
We refer to the analysis mimicking  multiband observations as  \emph{LISA+Earth}, and to the one using LISA alone as \emph{LISA-only}.
A proper multiband analysis will yield an even more significant impact on the parameter estimation, by providing valuable constraints on intrinsic parameters such as the mass ratio and the spins, which might be poorly constrained by observations with LISA only.  For this reason, our results in the \emph{LISA+Earth} case can be considered as conservative. 

In this paper, we consider only one GR modification at the time (either dipolar radiation or mass of graviton), and, as mentioned above, we neglect modifications to the amplitude of GWs.

\subsection{Signal generation}

We consider quasicircular binary systems consisting of spinning black holes with aligned or antialigned spins with respect to the orbital angular momentum. We omit possible orbital precession (which should be weak in the early inspiral) and orbital eccentricity in our model. Note that the eccentricity might not be negligible, in which case our analysis should be extended.  Each  system is characterised by 11 parameters: the masses ($m_1$ and $m_2$), the GW frequency at which LISA starts observing the system ($f_0$), the magnitude of the spins ($\chi_1$ and $\chi_2$), the position in the sky defined in the solar system barycenter ($\lambda$ and $\beta$), the polarisation angle ($\psi$), the azimuthal angle of the observer in the source frame ($\phi_0$), the inclination of the orbital angular momentum with respect to the line of sight ($\iota$, which is also the polar angle of an observer in the source frame) and the luminosity distance to the source ($D_L$). 

We consider three different astrophysical systems. System 1 one is similar to GW150914. Systems 2 and 3 were chosen to be significantly 
different from GW150914 to evaluate the dispersion of constraints/measurements of non-GR parameters between different systems. 
Since low mass systems are less likely to be detected by LISA, we opted for heavier systems. The distances were chosen to keep the SNR 
at a comparable level for all systems, so that it does not bias our results. Notice that System 3 has quite high spins, 
and as we will discuss in Sec. \ref{res} this can bias the measurement of deviations from GR.
In addition, for System 1 we consider three different values of the initial frequency, so that the time to coalescence 
(from the start of LISA observations) takes values of  $8.3 \ {\rm years}$, $4 \ {\rm years}$ and $2.5 \ {\rm years}$.  The choice of the initial 
GW frequency strongly affects the frequency evolution of the signal during  LISA observations, and allows us to explore constraints 
on GR modifications as a function of the signal's ``chirpiness''. 
The parameters of the  systems are given in Table \ref{params_syst}. We also provide the time to coalescence ($t_c$) and the SNR for each system, 
assuming $4 \  {\rm years}$ of LISA operation.

{\centering
  \begin{table}
   \begin{tabular}{|c|c|c|c|c|c|}
   \hline
   
    Variable & \multicolumn{3}{|c|}{System 1}  & System 2 & System 3 \\
  \hline \
  
  $m_1$ (M\textsubscript{\(\odot\)}) & \multicolumn{3}{|c|}{$38$} & $60$ & $50$ \\

  \hline
  
  $m_2$ (M\textsubscript{\(\odot\)}) & \multicolumn{3}{|c|}{$32$} & $50$ & $40$  \\
  
  \hline
    
$f_0$ (mHz) & \multicolumn{1}{|c|}{$12.4765$} & $16.4265$ & $19.5265$ & $12.3826$ & $12.783$ \\

\hline
  
  $\chi_1$ & \multicolumn{3}{|c|}{$0.05$} & $0.10$ &  $ 0.78$ \\
  
  \hline 
  
  $\chi_2$ &  \multicolumn{3}{|c|}{$0. 02$} & $0.33$ & $0.22$ \\
  
  \hline
  
  $\lambda$ (${\rm rad}$) & \multicolumn{3}{|c|}{$3.5064$}  & $0.2283$ & $2.9966$ \\
  
  \hline

   $\beta$ (${\rm rad}$) & \multicolumn{3}{|c|}{$0.1777$} & $-0.431$ & $-0.577$ \\
  
  \hline
  
   $\psi$ (${\rm rad}$) & \multicolumn{3}{|c|}{$1.1$} & $2.8$ & $1.5$ \\
  
  \hline
  
   $\phi_0$ (${\rm rad}$) & \multicolumn{3}{|c|}{$5.4$} & $6.0$ & $0.88$\\
  
  \hline
  
   $\iota$ (${\rm rad}$) &  \multicolumn{3}{|c|}{$2.77$} & $0.52$ & $0.34$ \\
  
  \hline
  
  $D_L$ (Mpc) & \multicolumn{3}{|c|}{$380$} & $640$ & $420$ \\
  
  \hline
  
  $t_c$ (yrs) &  \multicolumn{1}{|c|}{$8.3$} & $4$ & $2.5$ & $4$ & $ 5.2 $ \\
    
  \hline

  SNR &  \multicolumn{1}{|c|}{$10.5$} & $13.2$ & $10.9$ & $13.1$ & $15.0$ \\
  
  \hline

   \end{tabular}
    \caption{Systems considered in our analysis.}\label{params_syst}
  \end{table}}

We generate the frequency domain GW signal emitted by these systems as predicted by GR, i.e. $\tilde{h}_{GR}$ in Eq. (\ref{PPE}), by using PhenomD, a phenomenological waveform built from PN expressions and fits to numerical simulations. PhenomD  provides the dominant mode $\tilde{h}_{2\pm 2}$ \cite{Husa:2015iqa,Khan:2015jqa}. Modifications to GR are added to $\tilde{h}_{2\pm 2}$ as described in Sec. \ref{ppe_gw}.
We generate the signal between the initial frequency of the template, $f_0$, and the minimum between $0.5 \ {\rm Hz}$ (which we  assume to be LISA's Nyquist frequency) and the frequency reached by the end of observation (4 years). 

We compute the full LISA response to the GW signal using the method described in \cite{Marsat:2018oam}.
We work in the zero noise approximation in order to speed up the computation, but  we explore all possible correlations. Adding noise to the GW signal should not affect the parameter estimation drastically, leading mostly to a shift of the maximum likelihood within the quoted credible interval. Furthermore, the zero noise approximation could be seen as the average over many noise realisations \cite{Rodriguez:2013oaa}.

\subsection{Bayesian analysis}

We deploy a full Bayesian framework to explore the accuracy with which LISA can estimate the parameters of the source and put constraints on modifications to GR.
We treat all  parameters of the source, as well as the coefficients parametrising deviations from GR, as random variables, and we use Bayes' theorem to obtain the posterior distribution:
\begin{equation} \label{bayes} 
 p(\theta|d,M)=\frac{p(d|\theta, M)p(\theta|M)}{p(d|M)}.
 \end{equation}
 In this equation $p(\theta|d,M)$ is the posterior distribution that we want to sample, $p(d|\theta, M)$ is the likelihood, $p(\theta|M)$ is the prior distribution and $p(d|M)$ is the evidence. 
 
In our case the data $d$ corresponds to the injected signal, $\theta$ is the set of parameters characterising the signal, and $M$ is the model that we use to analyse the data. In more detail, the  models 
are GR, GR plus dipole radiation and GR plus  a massive graviton.

 Assuming the noise to be Gaussian, the likelihood is given by:
 \begin{equation}
  p(d|\theta, M)=e^{-\frac{1}{2}(d-h|d-h)},
 \end{equation}
where the inner product is defined as:
 \begin{equation} \label{correlation}
  (h_1|h_2)=4 \mathfrak{Re} \left(\int \frac{\tilde{h}_1(f) \ \tilde{h}_2^*(f)}{S_n(f)} {\rm d}f \right).
 \end{equation}
The denominator, $S_n(f)$, is the one-sided noise power spectral density (PSD). 
In this work we use the LISA ``Proposal'' PSD as specifically formulated in \cite{AHO17}. 
Details on the waveform generation and likelihood computation can be found in \cite{Marsat:2020rtl}. 
 
We take the priors to be flat in $m_1$ and $m_2$ with $m_1 \geq m_2$, flat in the spin magnitude ($\chi_1$ and $\chi_2$) between $-1$ and $1$, volume uniform for the distance to the source,  and isotropic for the sky position and inclination, while the polarisation and the observer's azimuthal angle priors are taken to be
uniform on the circle. In the \emph{LISA-only} scenario, we assume a flat prior for the initial frequency, whereas in the \emph{LISA+Earth} scenario we use a Gaussian prior with standard deviation $\sigma_{t_c}=1 \ {\rm ms}$ centred around the true value of the coalescence time $t_c$. 
Finally, we assume a flat prior for the coefficients parameterising deviations from GR.  We assume $B\geq 0$ and $m_g$ $\geq 0$, corresponding to positive extra GW fluxes (besides the GR ones) and  real positive masses respectively.

In order to sample the posterior distribution obtained through \eqref{bayes}, we use a Metropolis-Hashtings algorithm  
\cite{Karandikar2006,10.2307/2684568}. Its implementation is detailed in \cite{Toubiana:2020cqv} to which we refer for more details.
Instead of $m_1$, $m_2$, $\chi_1$, $\chi_2$, we use suitable combinations to explore the parameter space, namely the chirp mass ($\chirpm$), the symmetric mass ratio ($\eta$), the effective spin $\chi_+=\frac{m_1\chi_1+m_2\chi_2}{m_1+m_2}$ and the antisymmetric spin $\chi_-=\frac{m_1\chi_1-m_2\chi_2}{m_1+m_2}$. 
Instead of $m_g$, we use $m_{eff}=\frac{D_{0}m_g}{(1+z)D_L}$, in order to avoid the computation of $D_0$ (see Eq. \eqref{mod_disp}) at every point, thus saving computational time. Posteriors are reweighted at the end. 
Finally, it is noteworthy that when sampling the posterior, we allow the chains to explore negative values too and impose the cut a posteriori. This procedure reduces the number of effective points that we obtain out of a chain, but it allows for better sampling of the region close to the prior boundary.

In order to cross-check our results we also used the parallel tempering code {\bf PTMCMC} \cite{justin_ellis_2017_1037579}.
We obtained an excellent agreement between the two samplers, and especially for the $B_{{\rm BH}}$ and $m_g$ marginalised distributions.

\section{Results}\label{res}

We start by evaluating the constraints that we can place on the modifications of GR, and we then consider the problem of detecting those modifications 
(when present).

\subsection{Putting upper bounds on non-GR parameters}

 \begin{figure*}
\centering
 \includegraphics[width=0.8\textwidth]{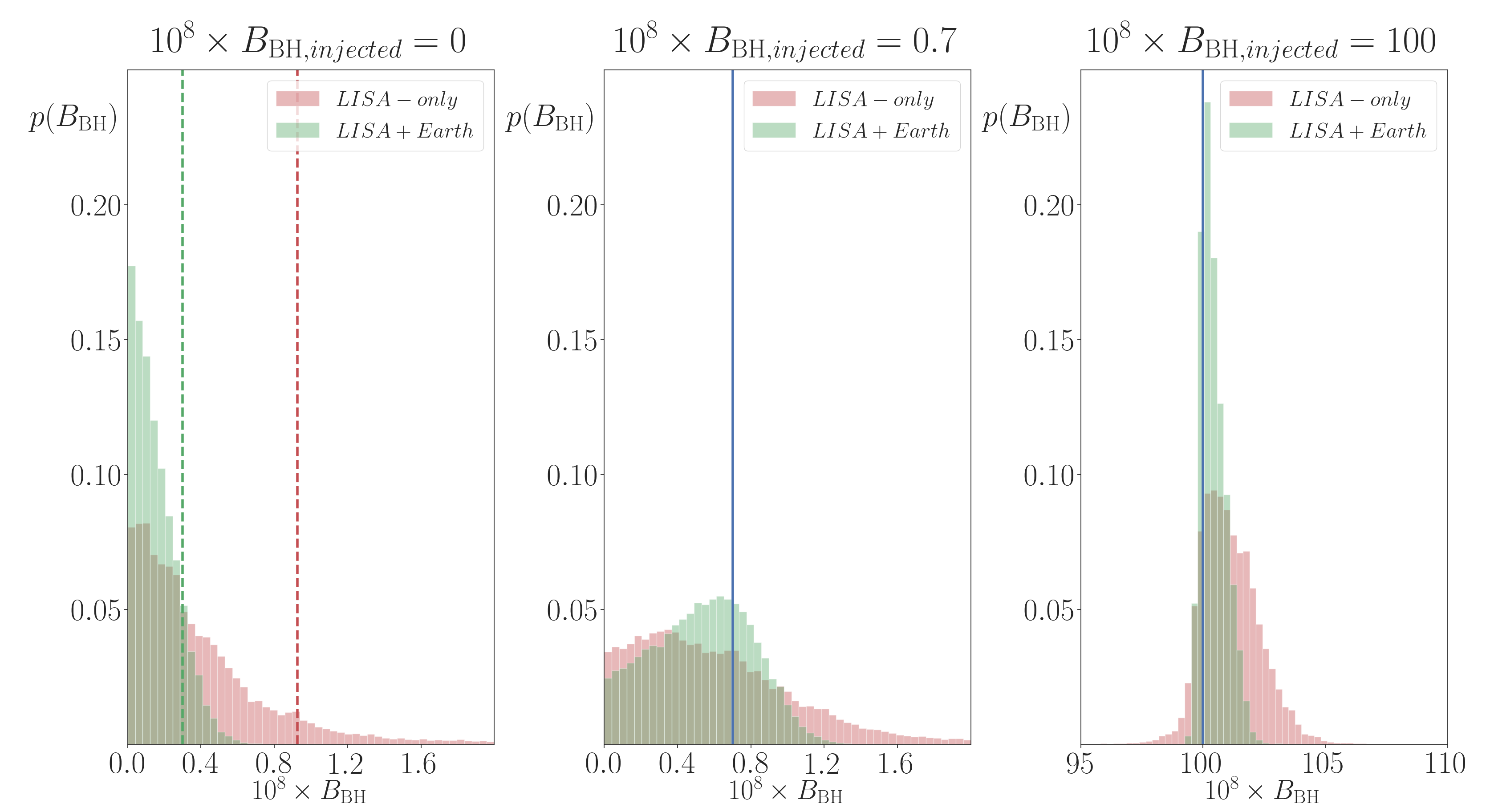}\\
 \centering
 \caption{Distribution of the dipolar amplitude for  System 1 (merging in 4 years), when using \emph{LISA-only} (red) and \emph{LISA+Earth} (green). In the left panel, the injected value is 0 and the dashed lines indicate the upper bound that we can put on B (corresponding to the $90 \% \ CI$). 
In both scenarios the upper bound is much below the current constraint ($4 \times 10^{-2}$)}. In the middle and right panels, the injected value is $0.7 \times 10^{-8}$ and $1.0 \times 10^{-6}$ respectively, indicated by the blue solid line. These values were chosen so that one is far above the bound that we can put on $B_{{\rm BH}}$ in the \emph{LISA+Earth} scenario, and the other is of the same order. Since 0 is not in the support of the posterior in the right panel, in this case we could safely claim  the detection of a modification to GR, unlike for the posterior in the middle panel.\label{dip_sys10}
\end{figure*}

We  assume that the GW signal follows GR (simulated data) and we use non-GR templates (waveform) as a search model for each  system given in Table \ref{params_syst}. The aim here is to set un upper limit on the phenomenological deviations from GR. 
In the left panel of Fig. \ref{dip_sys10} we show the marginalised distribution of $B_{{\rm BH}}$ for System 1 $(t_c=4.0 \ {\rm years})$ both in the \emph{LISA} and \emph{LISA+Earth} scenario. The distribution peaks at the true value of the dipolar amplitude (i.e. 0) and has a compact support extending up to the maximum $B_{{\rm BH}}$ compatible with observations. 

For most of the cases/systems considered here (with the exception of System 3, to which we will come back when discussing the possibility of detecting modifications to GR in \ref{section_det_mod}), we obtain similar distributions for $B_{{\rm BH}}$  and for $m_g$, as displayed in  the left panel of Fig. \ref{dip_sys10}. This allows us to place upper limits as 90\% credible interval (0.9 quantile of the corresponding marginalised distribution).  We present the upper bounds on $B_{{\rm BH}}$ and $m_g$ obtained with each system and in each scenario in Tables \ref{bounds_dip} and \ref{bounds_mg} respectively. We also provide the currently available  constraints  for comparison. 
The best constraints on dipole radiation in binary systems come from the binary pulsars (cf. Sec. \ref{ppe_gw}). However, since the value of the dipolar amplitude might depend on the nature of the system, as argued in Sec. \ref{ppe_gw}, we only 
consider here current bounds for systems containing at least one black hole. For those, the most stringent current bound comes from the observation of a low mass X-ray binary \cite{PhysRevD.86.081504}. This constraint is slightly better than the one obtained with current GW detections \cite{Barausse:2016eii}.
For $m_g$, we show the constraint obtained by the LIGO/VIRGO collaboration  during the first two observational runs~\cite{LIGOScientific:2019fpa}. 
A somewhat better upper bound on the mass of the graviton was obtained from solar system observations \cite{Will:2018gku}, but
it is unclear that a constraint from such a static configuration should be the same as for highly dynamical systems like black  hole binaries.

{
\renewcommand{\arraystretch}{1.5}
{\centering
  \begin{table} \label{constraints}
   \begin{tabular}{c|c|c|}
   
   \cline{2-3} 
   
   & \multicolumn{2}{|c|}{$B_{{\rm BH}}$} \\

   \cline{2-3}
   
   & \emph{LISA-only}  & \emph{LISA+Earth}   \\
   
   \hline
  
  \multicolumn{1}{|c|}{${\rm System} \ 1 (t_c=8.3 \ {\rm years})$} & $<1.1 \  10^{-7}$ & $<1.1 \ 10^{-7}$  \\

  \hline

  \multicolumn{1}{|c|}{${\rm System} \ 1 (t_c=4.0 \ {\rm years})$} & $<9.2 \  10^{-9}$ & $<3.2 \ 10^{-9}$  \\

  \hline

  \multicolumn{1}{|c|}{${\rm System} \ 1 (t_c=2.5 \ {\rm years})$} & $<6.8 \  10^{-8}$ & $<7.2 \ 10^{-9}$  \\

  \hline
  
 \multicolumn{1}{|c|}{${\rm System} \ 2$} & $<1.5 \  10^{-8}$ &  $<4.6 \ 10^{-9}$  \\

\hline
  
  \multicolumn{1}{|c|}{${\rm System} \ 3$} & $<1.9 \  10^{-7}$ & $<2.5 \ 10^{-8}$ \\
  
  \hline

  \multicolumn{1}{|c|}{Current constraints} & \multicolumn{2}{|c|}{$<4 \times 10^{-2}$}   \\

  \hline

   \end{tabular}
\caption{90 \% confidence constraints on $B_{{\rm BH}}$ obtained with each system in both the \emph{LISA-only} and \emph{LISA+Earth} scenario. Improvements by one of order of magnitude can be achieved when restricting $t_c$. Already in the \emph{LISA} scenario, all the systems considered in this work would allow one to improve current constraints for black hole systems.}\label{bounds_dip}
  \end{table}}
  }

{
\renewcommand{\arraystretch}{1.5}
{\centering
  \begin{table} \label{constraints2}
   \begin{tabular}{c|c|c|}
   
   \cline{2-3}
   
   & \multicolumn{2}{|c|}{$m_g \ ({\rm eV})$} \\
   
   \cline{2-3}
   
   & \emph{LISA-only}  & \emph{LISA+Earth}   \\
   
   \hline
  
  \multicolumn{1}{|c|}{${\rm System} \ 1 (t_c=8.3 \ {\rm years})$} & $<9.3 \  10^{-23}$ & $<2.5 \ 10^{-23}$  \\

  \hline

  \multicolumn{1}{|c|}{${\rm System} \ 1 (t_c=4.0 \ {\rm years})$} & $<2.0 \  10^{-23}$ & $<1.5 \ 10^{-23}$  \\

  \hline
  
  \multicolumn{1}{|c|}{${\rm System} \ 1 (t_c=2.5 \ {\rm years})$} & $<3.1 \  10^{-23}$ & $<2.5 \ 10^{-23}$  \\

  \hline
    
 \multicolumn{1}{|c|}{${\rm System} \ 2$} & $<1.2 \  10^{-23}$ &  $<1.2 \ 10^{-23}$  \\

\hline
  
  \multicolumn{1}{|c|}{${\rm System} \ 3$} & $<3.5 \  10^{-23}$ & $<2.0 \ 10^{-23}$ \\
  
  \hline

  \multicolumn{1}{|c|}{Current constraints} & \multicolumn{2}{|c|}{$<5 \times 10^{-23}$}   \\

  \hline

   \end{tabular}
\caption{90 \% confidence constraints on $m_g$ obtained with each system, in both the \emph{LISA-only} and \emph{LISA+Earth} scenario. As explained in the main text, restricting $t_c$ thanks to a multiband detection improves the bounds. Note that all the systems considered in this work would allow one to improve current constraints in the \emph{LISA+Earth} scenario.}\label{bounds_mg}
  \end{table}}
  }
  
The results presented in Tables \ref{bounds_dip} and \ref{bounds_mg} show that the best constraints are given by the systems which 
are observed during  the whole 4-year mission duration before passing out of band (System 2 and System 1 $(t_c=4.0 \ {\rm years})$).  In addition, System  1 $(t_c=2.5 \ {\rm years})$ gives better constraints than System 3 and System 1 $(t_c=8.3 \ {\rm years})$. Those results suggest that most of the constraining power comes from the chirp of the system. Although this might seem counterintuitive for low frequency modifications such as dipolar radiation, it can be explained by the correlation of non-GR parameters with intrinsic parameters such as mass ratio and spins. 
Indeed, we require substantial evolution of the signal in frequency to constrain those parameters even within GR, as discussed more extensively in \cite{Toubiana:2020cqv}.  Large uncertainties in $\eta$ and in the effective spin $\chi_+=\frac{m_1\chi_1+m_2\chi_2}{m_1+m_2}$ lead to larger errors on parameters correlated to them, such as $\chirpm$ and non-GR parameters. 


Overall, restricting $t_c$ improves the constraints on non-GR parameters, but it does not have the same impact for all systems. 
An interesting observation is that although System 3 gives a slightly worse constraint on $B_{{\rm BH}}$ than System 1 $(t_c=8.3 \ {\rm years})$
in the \emph{LISA-only} scenario, that constraint improves by an order of magnitude in the \emph{LISA+Earth} scenario, 
whereas the constraint from System 1 $(t_c=8.3 \ {\rm years})$ remains unchanged. To understand this, we transformed 
the samples obtained in the \emph{LISA-only} case to infer the time to coalescence of the systems, 
and we found that for all systems except System 1 $(t_c=8.3 \ {\rm years})$ there is a very 
strong correlation between $B_{{\rm BH}}$, $\chirpm$ and $t_c$. This is why restricting $t_c$ helps to improve the bound on $B_{{\rm BH}}$. 
Because we are observing System 1 $(t_c=8.3 \ {\rm years})$ at lower frequencies, the correlation between $B_{{\rm BH}}$ and $\chirpm$ is dominant. 
Thus restricting $t_c$ does not have much impact on the estimation of either parameter, and in particular the bound on $B_{{\rm BH}}$ does not improve. 
However, a real multiband detection would yield additional constraints on the parameters of the source and most likely improve the constraint on 
$B_{{\rm BH}}$ for System 1 $(t_c=8.3 \ {\rm years})$.

We see that the impact of restricting $t_c$  on the upper bound of $m_g$ is opposite: we get the tightest bound for the systems that start the
farthest from merger (System 1 $(t_c=8.3 \ {\rm years})$ and System 3). The reason is similar: LISA observes a GW signal 
at quite low frequency, while the effects of mass ratio, spins and $m_g$ appear beyond the leading PN order terms, and therefore they are 
(relatively) poorly constrained. Restricting $t_c$ to be in a narrow interval imposes an additional constraint on those parameters, which allows one 
to improve the bound on $m_g$. For systems closer to coalescence,  the frequency evolution during  LISA observations is sufficient to set tight bounds, and the additional constraint coming from restricting $t_c$ only moderately improves the results.

%

Our projected constraints on $B_{{\rm BH}}$ are in good agreement with \cite{Barausse:2016eii,Gnocchi:2019jzp} 
and should improve current bounds for BH systems by at most seven orders of magnitude. 
For $m_g$, we observe that multiband observations of any considered system should improve the current constraints by a factor few. Our bounds are somewhat better than the ones obtained with GW150914-like systems in \cite{Chamberlain:2017fjl,Carson:2019kkh}. 
One possible explanation for this difference is the use of sky averaged instrumental responses in theses works, while we use the full LISA response.
Indeed, as we discuss in \cite{Toubiana:2020cqv}, the use of approximations to the LISA response affects most the estimation of terms
appearing at higher frequencies, such as $m_g$.  
Finally, the reported upper bounds are worse than the projected bounds 
obtained from the observation of supermassive black holes binaries by LISA in \cite{Chamberlain:2017fjl}, due to a significant difference in 
the SNR and in the distance to those sources. We stress again that a real multiband analysis would not only 
restrict $t_c$, but also put additional constraints on all intrinsic parameters, improving  parameter estimation as a whole. 
As a consequence, bounds on non-GR parameters from SBHBs observed with LISA and ground detectors could be 
more stringent than those presented here in the \emph{LISA+Earth} scenario.

\subsection{Detecting modifications to general relativity}
\label{section_det_mod}
  
We now turn to the case where the injected signal has a nonzero value of either $B_{{\rm BH}}$ or $m_g$. 
Based on the results presented in Tables \ref{bounds_dip} and \ref{bounds_mg}, we choose $B_{{\rm BH}}$ well above the bounds presented there, but still below currents constraints ($B_{{\rm BH},injected}=100 \times 10^{-8}$), as well as $B_{{\rm BH}}$ and $m_g$ of the same order as those bounds ($B_{{\rm BH},injected}=0.7 \times 10^{-8}$ and $m_{g,injected}=1.0 \times 10^{-23} \ {\rm eV}$).  We choose a single value for $m_g$ because there is less room between our projected bounds and the currents constraints.
 
\begin{figure}
\centering
 \includegraphics[width=0.5\textwidth]{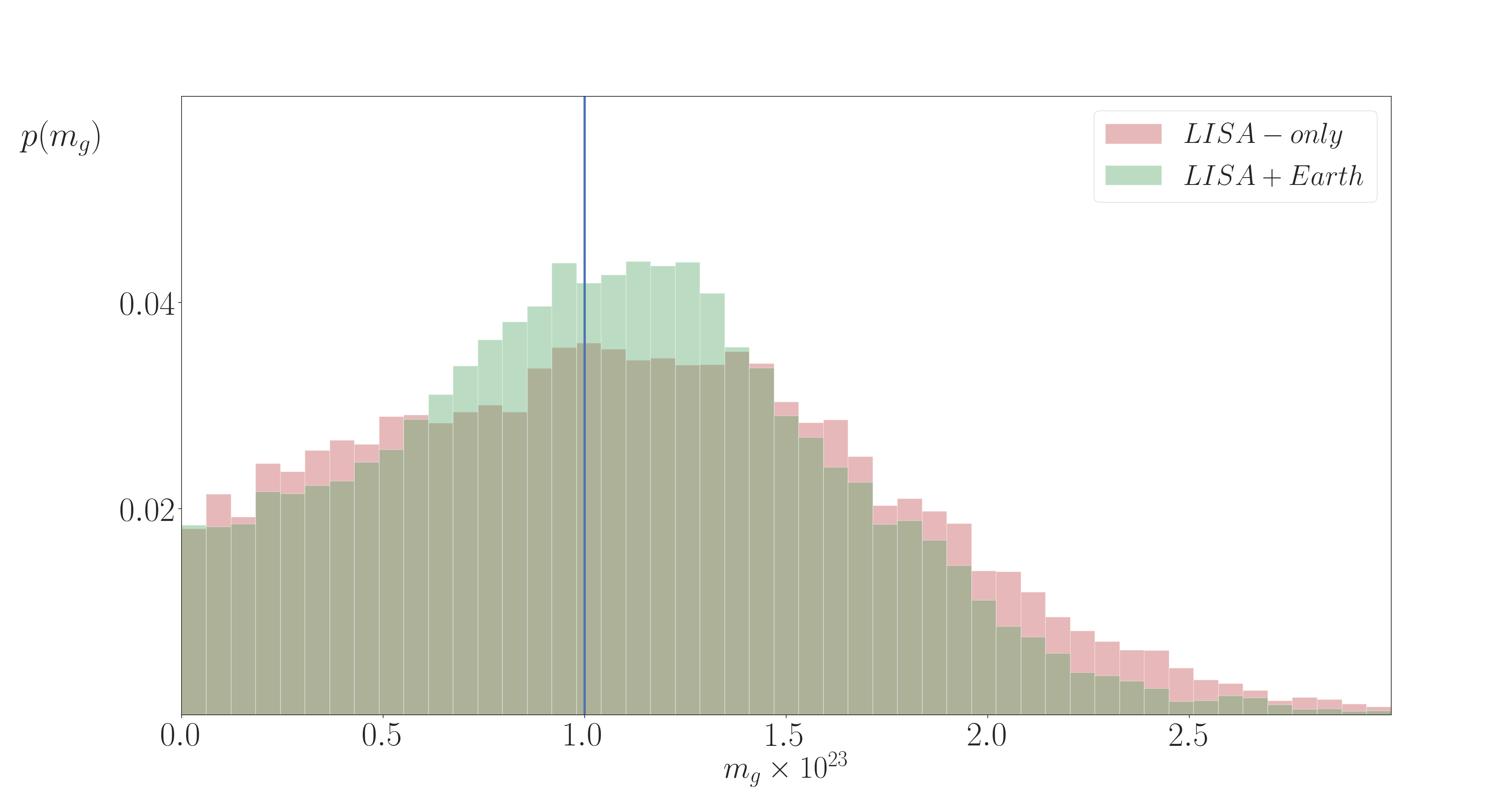}\\
 \centering
 \caption{Distribution of the mass of the graviton for System 2, when using \emph{LISA-only} (red) and \emph{LISA+Earth} (green). The injected value is $1 \times 10^{-23} \ {\rm eV}$, 
 indicated by the blue solid line, below the current constraint $5 \times 10^{-23}$. This value was chosen to be close to the upper bound that we can put in the \emph{LISA+Earth} scenario. Despite the peak around the injected value, 0 is in the support of the distribution, so we cannot safely claim the detection of a non-GR effect.}\label{mg_sys2}
\end{figure}

In the middle and right panels of Fig. \ref{dip_sys10}, we show the marginalised distribution of $B_{{\rm BH}}$ for System 1 $(t_c=4.0 \ {\rm years})$ both in the \emph{LISA} and \emph{LISA+Earth} scenario, when the injected value is $B_{{\rm BH},injected}=0.7 \times 10^{-9}$ and $B_{{\rm BH},injected}=100 \times 10^{-8}$ (respectively) denoted the blue solid line. For the higher $B_{{\rm BH}}$, the distribution peaks around the injected value and is not compatible with zero, clearly indicating the presence of the effect.
For the lower $B_{{\rm BH}}$, the distribution is very flat in the \emph{LISA-only} scenario and more peaked in the {\emph{LISA+Earth}} scenario. 
Similarly, in Fig. \ref{mg_sys2} we show the distribution of $m_g$ for System 2 and for an injected value of $10^{-23} \ {\rm eV}$. For the reasons explained in the previous section, the impact of restricting $t_c$ is milder, but it still slightly improves the sharpness of the posterior around the injected value.
Although the peak of the posterior distribution is away from zero (and centred on the true value),
as seen in Fig. \ref{dip_sys10}  and Fig. \ref{mg_sys2}  for the \emph{LISA+Earth} scenario, we  cannot still rule out GR (a vanishing deviation from GR  is compatible with the data), and we cannot safely claim the detection of a deviation from GR.

\begin{figure}
\centering
 \includegraphics[width=0.38\textwidth]{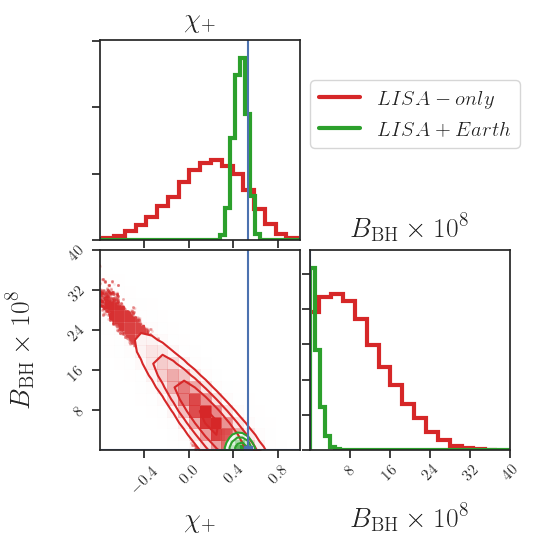}\\
 \centering
 \caption{Distribution of the dipolar amplitude for System 3, when using \emph{LISA-only} (red) and \emph{LISA+Earth} (green). The injected value is 0 and dashed lines indicate the upper bound that we can put on B (corresponding to the $90 \% \ CI$). Notice that in the \emph{LISA-only} case the distribution peaks away from 0, which would seem to indicate the presence of a nonzero modification, but this feature disappears in the \emph{LISA+Earth} case. This is due to the poor determination of  intrinsic parameters such as the spins, which leads to a bias in correlated parameters such as $B_{{\rm BH}}$. When constraining $t_c$, the determination of the intrinsic parameters improves, suppressing this bias.}\label{dip_sys3}
\end{figure}

%
Detailed analysis reveals that while analysing the GR signal, the peak of the distribution for $B_{{\rm BH}}$ could be away from zero even in the noise-free approximation. The corner plot  \cite{corner} presented in Fig. \ref{dip_sys3} shows the distributions of $\chi_+$ and $B_{{\rm BH}}$ and the correlation between them for System 3. Note that the true value of $B_{{\rm BH}}$ here is zero, however, in the \emph{LISA-only} scenario the distribution peaks at a non-zero value, mimicking a deviation from GR. Since we do not observe the chirp of this system, the likelihood is very shallow across the allowed range of  $\chi_+$, and the prior, which peaks at $\chi_+=0$, dominates. The high spins of the BHs entering  System 3 produce a rather high value of $\chi_+$, for which the prior has little support. Thus, we are biased in our estimate of $\chi_+$, and the strong correlation between $\chi_+$ and $B_{{\rm BH}}$ shifts the peak 
of posterior distribution for $B_{{\rm BH}}$ away from zero. In other words, our prior beliefs are stronger than the information (likelihood) provided by the data itself. Restricting $t_c$ puts some constraint on $\chi_+$, suppressing the bias in $B_{{\rm BH}}$. This bias for the \emph{LISA-only} scenario is much less obvious if the BHs spins are low 
like in  System 1.

{
\renewcommand{\arraystretch}{1.5}
  \begin{table*} 
  \centering
   \begin{tabular}{c|c|c|c|c|c|c|}
   \cline{2-7}
   
   & \multicolumn{2}{|c|}{$10^{8} \times B_{{\rm BH},injected}=0.7$} &  \multicolumn{2}{|c|}{$10^{8} \times B_{{\rm BH},injected}=100$} &  \multicolumn{2}{|c|}{$10^{23} \times m_{g,injected}=1.0$} \\
   
   \cline{2-7}
   
   & \emph{LISA-only}  & \emph{LISA+Earth}  & \emph{LISA-only}  & \emph{LISA+Earth} & \emph{LISA-only}  & \emph{LISA+Earth}  \\
   
   \hline
  
  \multicolumn{1}{|c|}{${\rm System} \ 1 (t_c=8.3 \ {\rm years})$} & $5.1\pm ^{+6.8}_{-5.1}$ & $5.1^{+6.8}_{-5.1}$ &  $105.0^{+15.5}_{-12.1}$ &  $102.8^{+12.8}_{-11.3}$ & $4.5\pm ^{+5.3}_{-4.5}$ & $1.7^{+2.0}_{-1.7} $ \\

  \hline

  \multicolumn{1}{|c|}{${\rm System} \ 1 (t_c=4.0 \ {\rm years})$} & $0.5^{+0.7}_{-0.5} $ & $0.54^{+0.3}_{-0.5}$ & $101.0^{+2.2}_{-1.4} $ & $100.3^{+1.1}_{-0.6}$ & $1.6^{+1.4}_{-1.6} $ & $1.1^{+1.0}_{-1.1} $   \\

  \hline

  \multicolumn{1}{|c|}{${\rm System} \ 1 (t_c=2.5 \ {\rm years})$} & $1.6^{+4.8}_{-1.6}$ & $0.6^{+0.6}_{-0.6}$  & $101.5^{+6.7}_{-6.4} $ & $100.6^{+2.0}_{-1.0}$ & $2.9^{+2.4}_{-2.9}  $ & $1.29^{+1.3}_{-1.3} $   \\

  \hline
  
  \multicolumn{1}{|c|}{${\rm System} \ 2$} & $0.7^{+1}_{-0.7}$ & $0.6^{+0.5}_{-0.6} $ & $101.1^{+2.8}_{-1.9}$ & $100.3^{+1.0}_{-0.6}$ & $1.1^{+0.9}_{-1.1}$ &  $1.1^{+0.7}_{-1.1} $  \\

\hline
  
  \multicolumn{1}{|c|}{${\rm System} \ 3$} & $8.8^{+11}_{-8.8}$ & $1.2^{+1.7}_{-1.2}  $ & $109.3^{+15.8}_{-11.8}$ & $103.4^{+4.6}_{-4.2}$ & $2.1^{+2.3}_{-2.1}$ & $1.9^{+2.0}_{-1.9}$ \\
  
  \hline

   \end{tabular}
\caption{Recovered $90\%$ CI for non-GR parameters for different injected non-zero values. For a strong modification $B=100 \times 10^8$ the posterior peaks around the injected value, and 0 is not in the support of the distribution, as in the right panel of Fig. \ref{dip_sys10}. Thus, we could safely claim the detection of a modification to GR. In the case of smaller modifications, e.g. $B_{{\rm BH},injected}=0.7 \times 10^8$ or $m_{g,injected}=1.0 \times 10^{-23} \ {\rm eV}$, the posterior is compatible with 0. Therefore, even in cases where the distribution peaks around the true value, like in Fig. \ref{mg_sys2} and in the middle panel of Fig \ref{dip_sys10}, we could not safely claim the detection of a non-GR effect.}\label{det_mod}
  \end{table*}
  }

  The exercise above indicates the importance of multiband observations. At the same time (especially in the presence of noise) we should be careful and claim the detection of a modification to GR only if the distribution of non-GR parameters is incompatible with zero. For that reason, whenever the posterior distribution is compatible with zero, we define the $90 \%$ confidence interval (CI) of $B_{{\rm BH}}$ and $m_g$ as the values between the 0 and the 0.9 quantiles. Otherwise, in situations as in the right panel of Fig. \ref{dip_sys10}, we define the CI as the values between the 0.05 and  0.95 quantiles, and we report the median as a point estimate.  The $90 \%$ CI on $B_{{\rm BH}}$ and $m_g$ are given for each case considered in Table \ref{det_mod}.

The errors on non-GR parameters in all systems and scenarios considered are coherent with upper bounds: the cases giving the more stringent bounds are the ones that could detect modifications with higher precision.

For completeness we add that, for systems where the injected value is much lower than the calculated upper bound (e.g. when $B_{{\rm BH},injected}=0.7 \times 10^{-8}$ for System 1 $(t_c=8.3 \ {\rm years})$ or System 3), we cannot distinguish a peak away from zero, and the distribution is very similar to the one in the left panel of Fig. \ref{dip_sys10}. 

In order to further investigate the presence of a modification to GR one could compute Bayes factors given by the ratio of evidences of two models : $\mathcal{B}_{M_1,M_2}=\frac{p(d|M_1)}{p(d|M_2)}$. As an example, Bayes factors much larger than 1 would suggest that model $M_1$ describes data better than model $M_2$. In our case the difference between models would be the presence (or absence) of some modification to GR in the GW templates. To carry this study we would need to use different samplers, e.g. nested 
sampling \cite{skilling2006}. Additionally, the evidences could be used to weigh the individual posteriors and combines obervations. 
Because the dipolar amplitude will in general be system dependent, stacking events in order to improve  the constraint on $B_{{\rm BH}}$ may not 
be meaningful. On the other hand, combined observations could definitely improve the constraint on the mass of the graviton \cite{LIGOScientific:2019fpa}. 
We leave these investigations for future work.

 \section{Conclusion}\label{ccl}
 
 In this paper we have investigated the possibility of  constraining modified theories of gravity through the observation of GWs from SBHBs with LISA.
 We have also performed a first assessment of the further improvements on these constraints that could be achieved  by multiband (LISA plus ground-based detectors) observations.
 In order to perform a theory agnostic estimation, we used the ppE framework, considering only leading order effects on the GW phase due to phenomenological modifications of GR. We have focused on two possible modifications of GR: (i) the existence of dipolar radiation, which affects the generation of GWs, and (ii) a nonzero mass of the graviton,  which affects the propagation of GWs.
 In order to be as realistic as possible, we simulated data using the phenomenological waveform PhenomD, taking into account the nontrivial response of LISA. We have chosen three fiducial systems consistent 
 with currently detected  LIGO/VIRGO binary black holes, and we explored the influence of initial 
 orbital frequency (or separation) on the constraints that could be placed on deviations from GR.
We have performed a full Bayesian analysis. Results obtained with a Metropolis-Hashtings Markov Chain Monte Carlo algorithm and a parallel tempering code {\bf PTMCMC} \cite{justin_ellis_2017_1037579} are in a very good agreement.
Overall, we find that multiband observations should improve current bounds on dipole radiation from black hole systems
 by seven  orders of magnitude, and current bounds on the  mass of the graviton by half an order of magnitude. 
 
We have identified several possible investigations as a continuation to this work. We have considered quasicircular orbits for simplicity, but it was shown in \cite{Porter:2010mb} that eccentricity could play an important role, and  we do not expect  all  SBHBs to have circularised before entering the LISA band \cite{Antonini:2012ad,Samsing:2017xmd}. 
 Furthermore, as discussed in \cite{Gerosa:2018wbw}, we might expect SBHBs  to have nonaligned spins.
Moreover, we have mimicked multiband observations by constraining the time to coalescence. However,
as discussed earlier in this paper, the contribution from a multiband detection does not consist solely in measuring the time to coalescence, as it should allow to put additional constraints on all the parameters of the source. We leave a more detailed investigation of the multiband analysis of SBHBs to future work. Finally, evaluation of the Bayes factor should be used for the model selection.

\section*{Acknowledgements}
E.B. acknowledges financial support provided under the European Union's H2020 ERC Consolidator Grant
``GRavity from Astrophysical to Microscopic Scales'' grant agreement no. GRAMS-815673.
 This work has also been supported by the European Union's Horizon 2020 research and innovation program under the
Marie Sk\l{}odowska-Curie grant agreement No 690904.
A.T. acknowledges financial support provided by Paris-Diderot University (now part of Université de Paris).
The authors would also like to acknowledge networking support from the COST
Action CA16104.  
 
 \FloatBarrier

 \bibliography{base_ref.bib}

\end{document}